\documentstyle[prl,aps,epsf,graphics,color]{revtex}

\newcommand{\Red}{\color{black}}
\newcommand{\Blue}{\color{black}}

\newcommand{\m}[1]{ {\Red $#1$} }

\newcommand{\beq}{\Red \begin{eqnarray}}
\newcommand{\eeq}{\nonumber\end{eqnarray}\Blue}
\newcommand{\numeq}{\end{eqnarray}\Blue}
\newcommand{\half}{ {1\over 2} }
\newcommand{\DOE}{This work was supported in part by U.S. Department of Energy
 grant No. DE--FG02-91ER40685}

\newcommand{\beqs}{\begin{eqnarray}}
\newcommand{\eeqs}{\end{eqnarray}}

\def\m#1{$#1$}

\begin{document}

\title{\bf\large The \m{x}-dependence of Parton Distributions Compared with Neutrino Data }

\author{
G. S. Krishnaswami\footnote{govind@pas.rochester.edu} and 
S. G. Rajeev\footnote{rajeev@pas.rochester.edu} \\
{\it Department of Physics and Astronomy, University of Rochester} \\
{\it Rochester, New York 14627}
}

\maketitle

\abstract{ We use the variational principle of Quantum HadronDynamics,
an alternative formulation of Quantum ChromoDynamics, to determine the
wavefunction of valence quarks in a baryon at a low value of
\m{Q^2}. This can be used to predict the structure function
\m{xF_3(x,Q^2)} at higher values of \m{Q^2} using the evolution
equations of perturbative QCD. This prediction is compared 
 to  the measurements of neutrino scattering
cross-section by the CDHS and CCFR experiments. The agreement is quite
good, confirming the validity of QHD as a way of studying hadronic structure.}

\vspace{.5 cm}



\vspace{.5 cm}

One of us has proposed \cite{2dqhd}  that it is possible to describe
 strong  interactions directly in terms of hadrons rather than in
 terms of quarks and gluons; this new approach was called
 Quantum HadronDynamics (QHD). It is  equivalent to Quantum
 ChromoDynamics (QCD, the universally accepted theory of strong
 interactions), except that the semi-classical approximation of QHD is
 a good approximation to nature: it is related to the large \m{N_c}
 limit of QCD.  Semi-classical methods applied directly to QCD give results
 that are too inaccurate except in the limit of short distances. QHD
 has been constructed so far only in two space-time
 dimensions. However, that is sufficient to obtain the structure
 functions of Deep Inelastic Scattering: in the limit of zero
 transverse momentum for the constituents, we can dimensionally reduce
 the theory to two space-time dimensions. More precisely, the 
 dependence of the structure functions on the Bjorken variable is
 determined by ignoring the transverse dimensions\cite{brodsky}. Thus we can predict the non-perturbative \m{x_B} dependence of structure functions, something that is completely inaccessible to the standard perturbative formulation of QCD. The 
leading order effects of 
transverse momenta are  described by the DGLAP \cite{dglap} equations
 that determine the \m{Q^2} dependence of the structure functions:
 this is the fundamental insight of the paper of Altarelli and Parisi.
 
In this paper we will focus on understanding the
 distribution of valence quarks in the nucleon. This is measured directly in neutrino
 (and anti-neutrino scattering) against the nucleon: the isospin
 averaged valence parton distribution function is just the structure
 function \m{F_3(x_B)} measured in this process \cite{cteq}.
To leading order in \m{1 \over \log(Q^2)}, it is not necessary to know the gluon structure functions of the
hadron in the DGLAP evolution of this quantity. The anti-quark
 content of the baryon is zero at the initial value \m{Q^2=Q_0^2}
 within our approximations. This is consistent with the phenomenological 
model of Gl\"uck and Reya\cite{grv}, that the anti-quark distribution 
of the proton is dynamically generated by \m{Q^2} evolution from an 
initial value of zero at low \m{Q^2}. We will, in a later paper\cite{antiquark}, 
calculate the anti-quark distribution functions and show that this is indeed  
justified. However, we expect the ``primordial'' gluon distribution to be non-zero.

In a previous paper \cite{qcdp} we derived  a variational principle that determines
this valence quark distribution function in the semi-classical
approximation of QHD. It was also shown there how to derive this
variational principle from QCD. In a separate paper \cite{ipm} we derived the
same principle from an interacting valence parton model of the
baryon. In this paper we will obtain approximate solutions of the
variational principle and compare  with the experimental data of the
CCFR and CDHS \cite{cdhsccfr} collaborations. The agreement of the predictions of QHD with data is quite spectacular: we now have a theory of the \m{x_B}
dependence of deep inelastic structure functions.

Let \m{\tilde\psi(p)} be the wavefunction of a valence parton in a
proton, thought of as a function of the null component of momentum 
(\m{p = p^0 - p^1}) \cite{2dqhd}. For kinematical reasons, \m{P>p>0},
where  \m{P} is the total momentum of the proton. Naively, we would
have the  sum rules,
\beq
	\int_0^P |\tilde\psi(p)|^2{dp\over 2\pi}=1,\quad
N_c\int_0^P p|\tilde\psi(p)|^2{dp\over 2\pi}=P.
\eeq
However, we have to modify the momentum sum rule since it is known
that the valence partons  carry only about half the momentum of the
proton: the rest is carried mostly by the gluons, with a small
contribution from anti-quarks. Thus we impose \m{N_c\int_0^P
p|\tilde\psi(p)|^2{dp\over 2\pi}=fP}, where \m{f} is the fraction of
the momentum carried by the valence partons. We will address the
problem of gluon distribution functions in a later paper, where we
will attempt to calculate \m{f}; here we will treat it as a parameter.

The wavefunction \m{\tilde\psi} is determined by minimizing the energy 
\beq
	E_1(\psi)=\int_0^P\half[p+{\mu^2\over
p}]|\tilde\psi(p)|^2{dp\over 2\pi}
+ {\tilde g^2\over 2}\int |\psi(x)|^2|\psi(y)|^2{|x-y|\over 2}dxdy
\eeq
subject to the above conditions. (Here \m{\psi(x)=\int_{0}^{P}
\tilde\psi(p)e^{ipx}{dp\over 2\pi}} is the wave-function in
position space.) This variational principle was
derived from QCD through QHD in Ref. \cite{qcdp} as well as from an
interacting parton model in Ref. \cite{ipm}. It describes 
partons (within the mean field approximation) which are interacting
with each other through a linear Coulomb potential, which binds them
into a baryon.  The linear potential comes from eliminating the gluon
fields: it is the one-dimensional Fourier transform of the gluon
propagator.

 A Lorentz invariant
formulation  (which is more convenient for our purposes) would be to
minimize the mass$^2$  of the  baryon rather than its energy:
\beq
	{\cal M}^2=\bigg[\int {p\over 2}|\tilde\psi(p)|^2{dp\over 2\pi}\bigg]\bigg[
\int_0^P{\mu^2\over 2p}|\tilde\psi(p)|^2{dp\over 2\pi} + {\tilde g^2 \over 2}\int
|\psi(x)|^2|\psi(y)|^2{|x-y| \over 2}dxdy\bigg].
\eeq
We are ignoring the flavor and spin  quantum numbers of  the partons
so what we get will be the spin and flavor averaged wavefunction. Also
\m{\mu^2~=~m^2~-~{\tilde g^2\over \pi}} is the quark mass$^2$ after a finite
 renormalization \cite{hagen}, and m is the current quark mass. The dimensional 
parameter \m{\tilde g \sim \Lambda_{QCD}} determines the strength of the interaction. 
We now define the number  density of valence partons 
\beq
	V(x_B)= N_c\left[1+C_1({\alpha_s(Q_0^2)\over \pi}) + C_2({\alpha_s(Q_0^2)\over\pi})^2 + C_3({\alpha_s(Q_0^2)\over\pi})^3 \right]{P\over 2\pi}
\left|\tilde\psi(x_BP)\right|^2
\eeq
We have normalized this so that the Gross-Lewellyn-Smith 
sum rule\cite{gls} (including the perturbative corrections 
up to order \m{\alpha_s^3(Q_0^2)}) is satisfied. 
The coefficients \m{C_1, C_2} and \m{C_3} are given in 
Ref.\cite{gls}. We will finally set the number of colors 
\m{N_c} to 3 and the number of flavours to 2 at 
the initial low value of \m{Q_0^2}.

Since the total momentum scales like 
\m{P\sim N_c}, in the limit as \m{N_c\to \infty}, the parton momentum has
the range \m{0 \leq p<\infty}. In the limit when  \m{N_c\to \infty} and \m{ m=0},
  we have found an {\it exact}  minimum of the 
variational principle: \m{\tilde\psi(p)=Ce^{-{p\over g}}.} 
The condition for minimizing \m{{\cal M}^2} is an integral equation and we
can verify that this function  is a solution by explicit computation.
 (The calculation   involves infrared singular integrals,
which are  defined through an appropriate
principal value prescription as in \cite{ipm}.)
This suggests that even when \m{N_c} is finite, a reasonable variational
ansatz would be \m{\tilde\psi(p)=C\left({p\over g}\right)^a[1-{p\over P}]^b.}
(Recall that \m{e^{-x}=\lim_{n\to\infty}[1-{x\over n}]^n}.) The
constant \m{C} is determined by the  normalization condition. 
By using the momentum sum rule, we get \m{b={N_c\over 2f}-1+a[{N_c\over f}-1].}
``\m{a}'' is determined by the variational principle. It is 
zero in the limit of chiral symmetry and rises like a power of 
\m{{m^2\over\tilde{g}^2}}. In the limit \m{N_c \to \infty}, 
\m{a} is determined by the transcendental equation
\beq
{\pi m^2\over \tilde g^2}=1+\int_0^1{dy\over y^2}\bigg[(1+y)^{a}+ (1-y)^{a}-2\bigg]+
\int_1^\infty {dy\over y^2}\bigg[(1+y)^a-2\bigg]
\eeq
which we derived by a Frobenius-type analysis of the integral equation 
for the minimization of \m{{\cal M}^2} in an earlier paper\cite{ipm}. 
Thus in the limit \m{m = 0}, \m{a = 0}, and we have the valence parton 
distribution function \m{V(x_B)=C[1-x_b]^{{N_c\over f}-2}}
where \m{C} is fixed by the GLS sum rule. 
This variational approximation agrees well with our numerical 
solution of the same problem in Ref. \cite{ipm} but is much simpler to use. 
The limit of chiral symmetry is a good approximation in the case of 
the nucleon since the current quark masses of the up and 
down quarks (5-8 MeV) are small compared to the energy scale 
of the strong interactions (\m{\Lambda_{QCD} \sim} 200 MeV).
The valence parton distribution depends on \m{\tilde{g}} only through 
the ratio \m{m^2\over\tilde{g}^2}
and in the limit \m{m = 0}, is independent of \m{\tilde{g}}. 

The structure function \m{F_3(x_B,Q^2)} of neutrino scattering on an
isoscalar target has been accurately measured by the CCFR and CDHS\cite{cdhsccfr}
experimental groups at several  values of \m{x_B} and \m{Q^2}. 
To compare with that data we need to evolve our computed distribution
function to the appropriate values of \m{Q^2} by the DGLAP equation:
\beq
	{dV(x_B,t) \over dt} = {\alpha_s(t) \over 2\pi} 
\int_{x_B}^{1} {dy \over y} V(y,t) P_{qq}({x_B \over y}),
\eeq
where \m{t = \log(Q^2/Q_0^2)}, \m{\alpha_s(t)} is evaluated 
using the two-loop \m{\beta} function\cite{buraspqcd} and \m{P_{qq}} 
is the evolution kernel to leading order given in Ref. \cite{dglap}.
We solve the evolution equation numerically, assuming an initial value of
\m{Q_0^2~=~0.4\;{\rm GeV}^2}. This low value of \m{Q_0^2} is justified
since we  can show\cite{antiquark} that the anti-quark
distribution function of the nucleon is quite small. The range of \m{Q}
over which we are evolving is small (\m{Q \sim 0.6} to \m{5~GeV}) and its effect is 
small, thus justifying the use of the leading order DGLAP equation. We also set 
\m{N_c=3}, \m{\Lambda_{QCD}=200} MeV and the current quark mass \m{m = 0}. As for
the fraction of baryon momentum carried by valence partons \m{f}, 
we choose it to be \m{0.5}, which is in agreement with the
phenomenological fits of parton distribution functions ( See
e.g.,\cite{pdfs}.) In a later paper we will derive the gluon structure
function as well, and at that time we will have a theoretical prediction for this parameter.
The plots show that we have quite good agreement with data. Thus we
have established that Quantum HadronDynamics is a successful way of
deriving hadronic structure functions from QCD.

Acknowledgements: We thank A. Bodek and especially V. John for discussions. 
\DOE.

\begin{center}
\scalebox{0.5}{\includegraphics{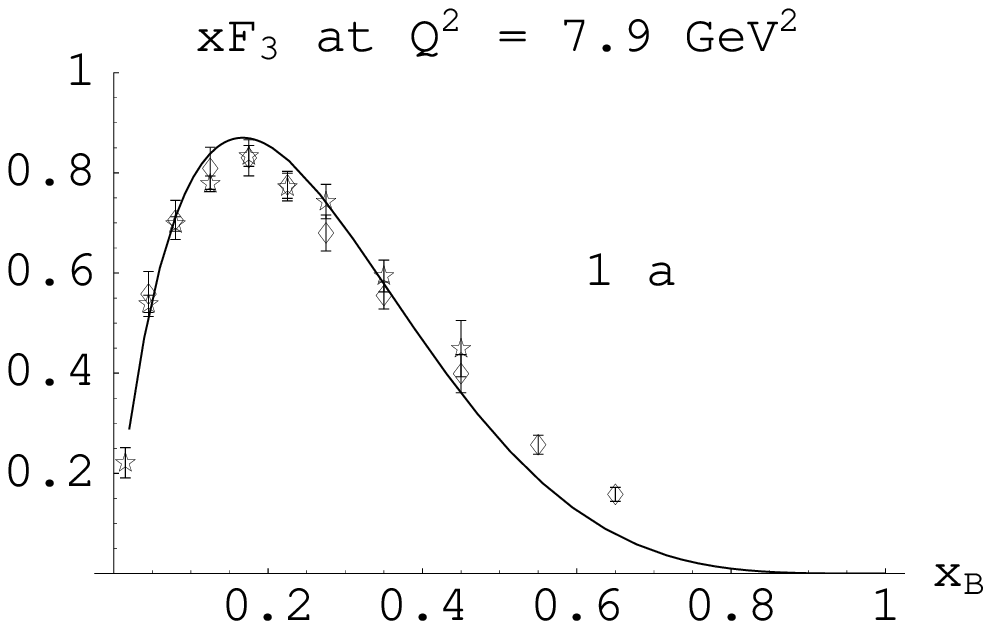}}
\scalebox{0.5}{\includegraphics{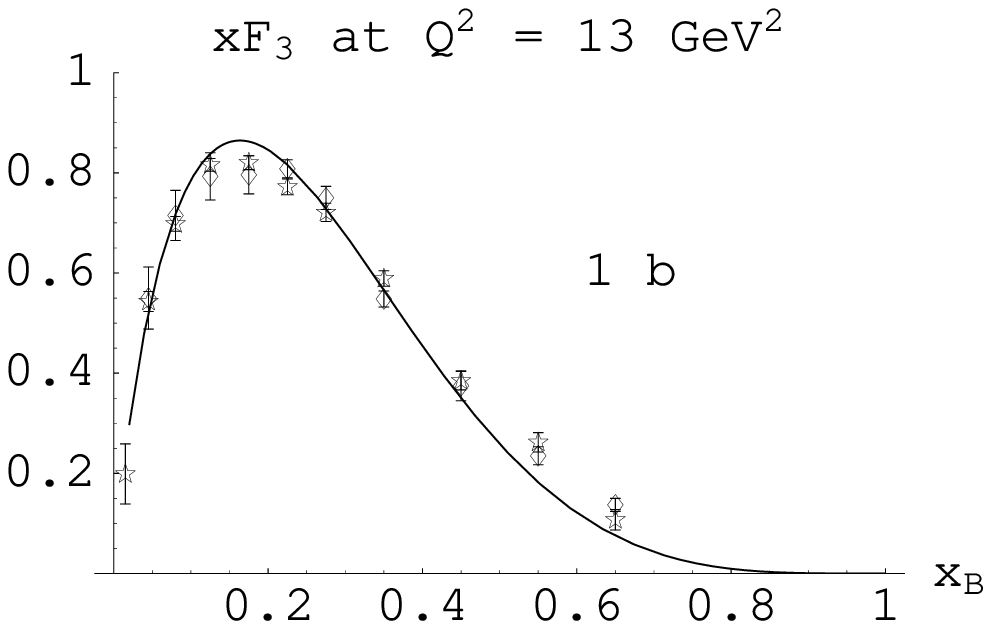}}
\end{center}

Figure 1: Comparison of QHD prediction of \m{xF_3} (solid curve) 
with measurements by CCFR(\m{\star}) and CDHS(\m{\diamondsuit}). 
We have chosen the parameters \m{Q_0^2~=~0.4~GeV^2} and \m{f~=~0.5}. 
(a) CCFR at \m{Q^2~=~7.9~GeV^2}, CDHS at \m{7.13 \leq~Q^2~\leq~8.46~GeV^2} 
and QHD prediction at \m{Q^2~=~7.9~GeV^2}. (b) CCFR at \m{Q^2~=~12.6~GeV^2}, 
CDHS at \m{12.05~\leq~Q^2~\leq~14.3~GeV^2} and QHD prediction at \m{Q^2~=~13~GeV^2}.

\begin{center}
\scalebox{0.5}{\includegraphics{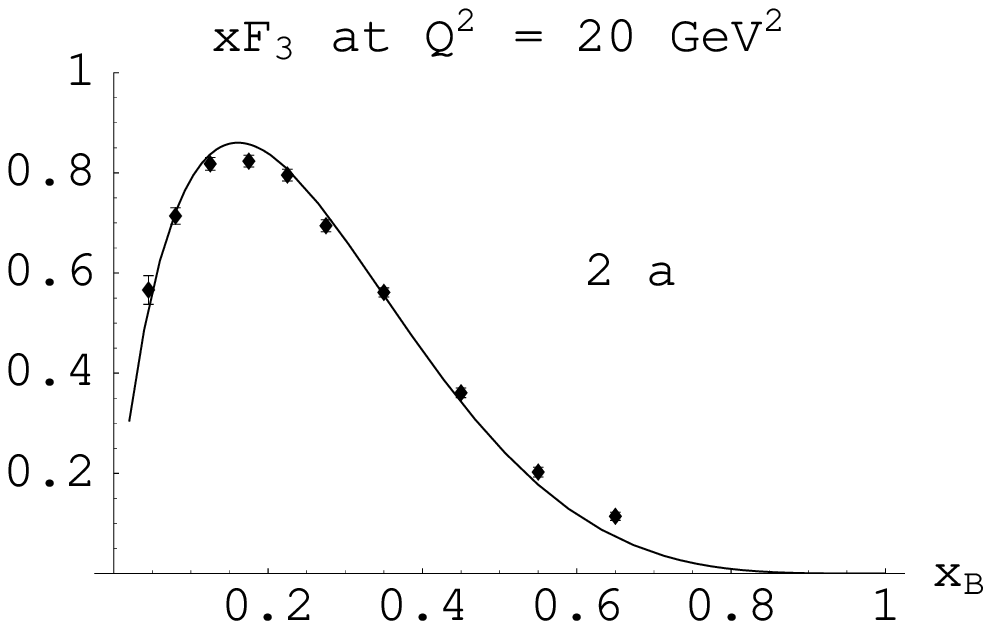}}
\scalebox{0.5}{\includegraphics{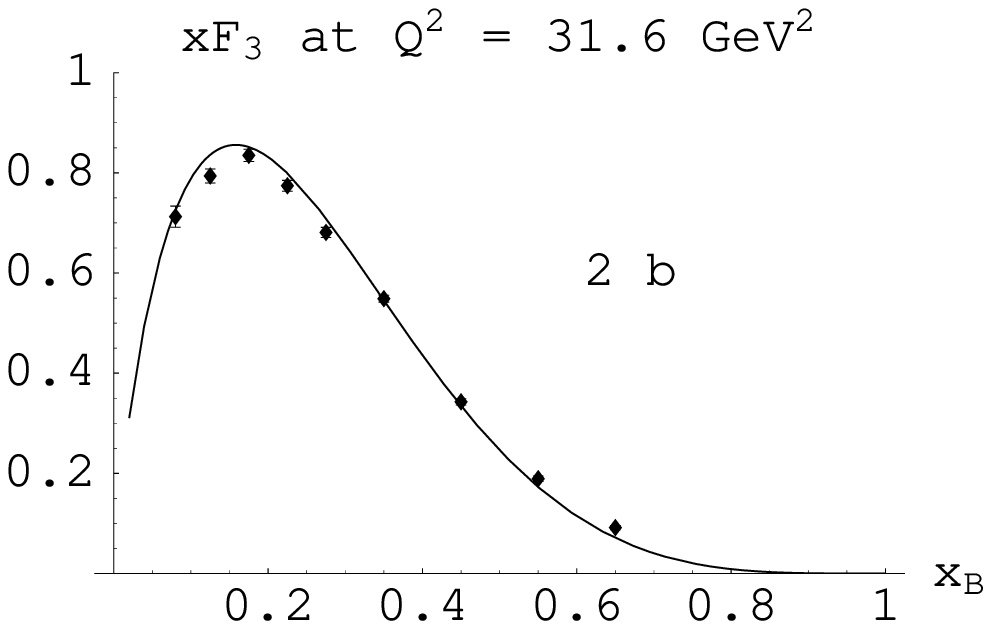}}
\end{center}

Figure 2: Comparison of QHD prediction of \m{xF_3} (solid curve)
with CCFR(\m{\bullet}) measurements at (a) \m{Q^2~=~20~GeV^2} 
and (b) \m{Q^2~=~31.6~GeV^2}.

\end{document}